\documentclass[aps, showpacs, showkeys,nofootinbib,floatfix]{revtex4}

\usepackage{amssymb}
\usepackage{amsmath}
\usepackage{graphicx}

\begin{document}

\title{Cosmic Constraint on Ricci Dark Energy Model}

\author{Lixin Xu\footnote{Corresponding author}}
\email{lxxu@dlut.edu.cn}
\author{Wenbo Li}
\author{Jianbo Lu}
\author{Baorong Chang}

\affiliation{Institute of Theoretical Physics, School of Physics \&
Optoelectronic Technology, Dalian University of Technology, Dalian,
116024, P. R. China}

\begin{abstract}
In this paper, a holographic dark energy model, dubbed Ricci dark
energy, is confronted with cosmological observational data from type
Ia supernovae (SN Ia), baryon acoustic oscillations (BAO) and cosmic
microwave background (CMB). By using maximum likelihood method, it
is found out that Ricci dark energy model is a viable candidate of
dark energy model with the  best fit parameters:
$\Omega_{m0}=0.34\pm 0.04$, $\alpha=0.38\pm 0.03$ with $1\sigma$
error. Here, $\alpha$ is a dimensionless parameter related with
Ricci dark energy $\rho_{R}$ and Ricci scalar $R$, i.e.,
$\rho_{R}\propto \alpha R$.
\end{abstract}


 \hfill ITP-DUT/2008-08

\maketitle

The observation of the Supernovae of type Ia
\cite{ref:Riess98,ref:Perlmuter99} provides the evidence that the
universe is undergoing accelerated expansion. Jointing the
observations from Cosmic Background Radiation
\cite{ref:Spergel03,ref:Spergel06} and SDSS
\cite{ref:Tegmark1,ref:Tegmark2}, one concludes that the universe at
present is dominated by $70\%$ exotic component, dubbed dark energy,
which has negative pressure and pushes the universe to accelerated
expansion. Of course, the accelerated expansion can attribute to the
cosmological constant naturally. In fact, the mostly observational
data is roughly consistent with the predictions of the $\Lambda$CDM
model \cite{ref:LCDM1,ref:LCDM2,ref:LCDM3}. However, it suffers the
so-called {\it fine tuning} and {\it cosmic coincidence} problem. To
avoid these problems, dynamic dark energy models are considered,
such as quintessence
\cite{ref:quintessence01,ref:quintessence02,ref:quintessence1,ref:quintessence2,ref:quintessence3,ref:quintessence4},
phtantom \cite{ref:phantom}, quintom \cite{ref:quintom} and
holographic dark energy \cite{ref:holo1,ref:holo2} etc. The
holographic dark energy is considered as a dynamic vacuum energy. It
is constructed by considering the holographic principle and some
features of quantum gravity theory. According to the holographic
principle, the number of degrees of freedom in a bounded system
should be finite and has relations with the area of its boundary. By
applying the principle to cosmology, one can obtain the upper bound
of the entropy contained in the universe. For a system with size $L$
and UV cut-off $\Lambda$ without decaying into a black hole, it is
required that the total energy in a region of size $L$ should not
exceed the mass of a black hole of the same size, thus
$L^3\rho_{\Lambda} \le L M^2_{pl}$. The largest $L$ allowed is the
one saturating this inequality, thus $\rho_{\Lambda} =3c^2
M^{2}_{pl} L^{-2}$, where $c$ is a numerical constant and $M_{pl}$
is the reduced Planck Mass $M^{-2}_{pl}=8 \pi G$.
It just means a {\it duality} between UV cut-off and IR cut-off. The
UV cut-off is related to the vacuum energy, and IR cut-off is
related to the large scale of the universe, for example Hubble
horizon, event horizon or particle horizon as discussed by
\cite{ref:holo1,ref:holo2}. In the paper \cite{ref:holo2}, the
author took the future event horizon
\begin{equation}
R_{eh}(a)=a\int^{\infty}_{t}\frac{dt^{'}}{a(t^{'})}=a\int^{\infty}_{a}\frac{da^{'}}{Ha^{'2}}
\end{equation}
as the IR cut-off $L$. As pointed out in \cite{ref:holo2}, it can
reveal the dynamic nature of the vacuum energy and provide a
solution to the {\it fine tuning} and {\it cosmic coincidence}
problem. In this model, the value of parameter $c$ determines the
property of holographic dark energy. When $c\ge 1$, $c=1$ and $c\le
1$, the holographic dark energy behaviors like quintessence,
cosmological constant and phantom respectively.

Recently, Gao, {\it et. al.} took the Ricci scalar as the IR cut-off
and named it Ricci dark energy \cite{ref:Gao}. In that paper
\cite{ref:Gao}, it has shown that this model can avoid the causality
problem and naturally solve the coincidence problem of dark energy.
In this paper, we will revisit the Ricci dark energy model and use
cosmic observations to constrain the model parameters. At first, we
give a brief review of the Ricci dark energy (RDE). We consider a
Friedmann-Robertson-Walker universe filled with cold dark matter and
RDE. Its metric is written as
\begin{equation}
ds^2=-dt^2+a^2(t)\left(\frac{dr^2}{1-kr^2}+r^2d\theta^2+r^2\sin^2\theta
d\phi^2\right),
\end{equation}
where $k=1,0,-1$ for closed, flat and open geometries respectively.
The Friedmann equation is
\begin{equation}
H^2=\frac{8\pi
G}{3}\left(\rho_{m}+\rho_{R}\right)-\frac{k}{a^2},\label{eq:FE}
\end{equation}
where $H$ is the Hubble parameter, $\rho_{m}$ and $\rho_{R}$ denote
the energy density of cold dark matter and Ricci dark energy
respectively. As suggested by Gao {\it et. al.}, the RDE is
proportional to the Ricci scalar
\begin{equation}
R=-6\left(\dot{H}+2H^2+\frac{k}{a^2}\right).
\end{equation}
Then, it is given as
\begin{equation}
\rho_{R}=\frac{3\alpha}{8\pi G}\left(\dot{H}+2H^2+\frac{
k}{a^2}\right)\propto R,\label{eq:rhoRDE}
\end{equation}
where $\alpha$ is dimensionless model parameter which can be
determined by confronting with cosmic observations. Now, the
Friedmann equation (\ref{eq:FE}) can be rewritten as
\begin{equation}
H^2=\frac{8\pi G}{3}\rho_{m0}{e^{-3x}}+\left(\alpha
-1\right)ke^{-2x}+\alpha\left(\frac{1}{2}\frac{dH^2}{dx}+2H^2\right)
\end{equation}
where $x=\ln a$. Dividing above equation by $H^2_0$ in both sides,
one has
\begin{equation}
E^2=\left(1-\alpha\right)\Omega_{k} e^{-2x}+\Omega_{m0} e^{-3x}
+\alpha\left(\frac{1}{2}\frac{dE^2}{dx}+2E^2\right), \label{eq:E}
\end{equation}
where $E\equiv H/H_0$, $\Omega_{k}=-k/H^2_0$ and
$\Omega_{m0}=\frac{8\pi G \rho_{m0}}{3H^2_0}$ are adopted. Solving
this equation, one has
\begin{eqnarray}
E^2&=&\Omega_{k}
e^{-2x}+\Omega_{m0}e^{-3x}+\frac{\alpha}{2-\alpha}\Omega_{m0}e^{-3x}+f_0e^{-\left(4-\frac{2}{\alpha}\right)x}\nonumber\\
&=&\Omega_{k}
e^{-2x}+\Omega_{m0}e^{-3x}+\Omega_{R}(x),\nonumber\\
&=&\Omega_{k}(1+z)^2+\Omega_{m0}(1+z)^3+\Omega_{R}(z)\label{eq:E}
\end{eqnarray}
where $f_0$ is the integral constant which is determined by the
current values of cosmological parameters, $\Omega_{R}$ is defined
as the dimensionless energy density of RDE in terms of $x$ or
redshift $z$
\begin{eqnarray}
\Omega_{R}(x)&=&\frac{\alpha}{2-\alpha}\Omega_{m0}e^{-3x}+f_0e^{-\left(4-\frac{2}{\alpha}\right)x}\nonumber\\
&=&\frac{\alpha}{2-\alpha}\Omega_{m0}(1+z)^3+(1-\Omega_k-\frac{2}{2-\alpha}\Omega_{m0})(1+z)^{\left(4-\frac{2}{\alpha}\right)}.\label{eq:omegaR}
\end{eqnarray}
In the second equal sign of the above equation, the initial
condition $E_0=1$, i.e. $\Omega_{k}+\Omega_{m0}+\Omega_{R0}=1$ is
used. Its equivalent $f_0=1-\Omega_k-\frac{2}{2-\alpha}\Omega_{m0}$
is given.

Considering the conservation equation of RDE, one has
\begin{equation}
w_{R}=-1-\frac{1}{3}\frac{d\ln\Omega_{R}}{dx}=-1+\frac{(1+z)}{3}\frac{d\ln\Omega_{R}}{dz}.\label{eq:wR}
\end{equation}
Now, we give a discussion on RDE model. In the above review, the
case of $\alpha=2$ is not taken into accounts. When $\alpha=2$, one
has the solution of Eq. (\ref{eq:E}) in terms of $x$ and redshift
$z$ as follows
\begin{eqnarray}
E^2&=&f_0e^{-3x}+\Omega_{k}e^{-2x}-\Omega_{m0}xe^{-3x}\nonumber\\
&=&f_0(1+z)^3+\Omega_{k}(1+z)^2+\Omega_{m0}(1+z)^3\ln(1+z)\nonumber\\
&=&\Omega_{m0}(1+z)^3+\Omega_{k}(1+z)^2+\Omega_{R}(z),
\end{eqnarray}
where $f_0$ is an integral constant. In the third equal sign of
above equation, noticing the first term behaves like cold dark
matter, we let $f_0=\Omega_{m0}$ and defined
$\Omega_{R}(z)=\Omega_{m0}(1+z)^3\ln(1+z)$ as RDE. While considering
the current values of cosmological parameters, one finds that
$\Omega_{R0}=0$ which is not consistent with the cosmic
observations. Then, the case of $\alpha=2$ is not a viable dark
energy model. Once one has $\alpha\ne 2$, it can be seen from Eq.
(\ref{eq:omegaR}) that the RDE contains two terms: the first one
behaves like cold dark matter, and the second one behaves like
exotic energy component which properties are determined by the
parameter $\alpha$. For its importance of the parameter $\alpha$, in
this paper, we will confront this model with current cosmic
observations including type Ia supernovae (SN Ia), baryon acoustic
oscillations (BAO) and cosmic microwave background (CMB) to test its
viability.

The maximum likelihood method will be used to constrain the RDE
model. The SN Ia data used in this paper contains $192$ SN Ia data
\cite{ref:sn192} compiled from the ESSENCE~\cite{essence} and Gold
sets \cite{Riess:2004nr,Astier:2005qq,Riess:2006fw}. Constraints
from Sne Ia can be obtained by fitting the distance modulus $\mu(z)$
\begin{equation}
\mu_{th}(z)=5\log_{10}(D_{L}(z))+\mathcal{M},
\end{equation}
where, $D_{L}(z)$ is the Hubble free luminosity distance $H_0
d_L(z)/c$ and
\begin{eqnarray}
d_L(z)&=&c(1+z)\int_{0}^{z}\frac{dz^{\prime}}{H(z^{\prime})}\\
\mathcal{M}&=&M+5\log_{10}\left(\frac{cH_{0}^{-1}}{Mpc}\right)+25
\nonumber\\
&=&M-5\log_{10}h+42.38,
\end{eqnarray}
where, $M$ is the absolute magnitude of the object (SN Ia here), and
$H_0$ is the Hubble constant which is denoted in a renormalized
quantity $h$ defined as $H_0 =100 h~{\rm km ~s}^{-1} {\rm
Mpc}^{-1}$. For SN Ia dataset, the best fit values of parameters in
a model can be determined by minimizing
\begin{equation}
\chi_{SNIa}^2(p_s)=\sum_{i=1}^{N}\frac{\left(\mu_{obs}(z_i)-\mu_{th}(p_s;z_i)\right)^2}{\sigma^2_{i}},\label{eq:chi2SNIa}
\end{equation}
where $N=192$ for the combined SN Ia dataset, $\mu_{obs}(z_i)$ is
the distance moduli obtained from observations, $\sigma_{i}$ is the
total uncertainty of the SN Ia data, and $p_s$ denotes the
parameters contained in the model.

The size of Baryon Acoustic Oscillation (BAO) is found by Eisenstein
{\it et al} \cite{ref:Eisenstein05} by using a large spectroscopic
sample of luminous red galaxy from SDSS and obtained a parameter $A$
which does not depend on dark energy models, in a flat universe.
And, it is written as
\begin{equation}
A=\frac{\sqrt{\Omega_{m0}}}{E(z_{bao})^{1/3}}\left[\frac{1}{z_{bao}}\int_{0}^{z_{bao}}\frac{dz}{E(z)}\right]^{2/3},
\end{equation}
where, $z_{bao}=0.35$ and $A=0.469\pm0.017$. The parameter has been
used to constrain the dark energy models. From BAO, the best fit
values of parameters in dark energy models can be determined by
minimizing
\begin{equation}
\chi^{2}_{BAO}(p_s)=\frac{(A(p_s)-0.469)^2}{0.017^2}.\label{eq:chi2BAO}
\end{equation}

The constraint to dark energy from CMB can be used is the CMB shift
parameter $R$ \cite{ref:Bond1997},
\begin{equation}
R=\sqrt{\Omega_{m0}}\int_{0}^{z_{rec}}\frac{dz}{E(z)},
\end{equation}
where $z_{rec}=1089$ is the redshift at the last scattering surface.
The $R$ obtained from $3$-year WMAP data \cite{ref:Spergel06} is
\cite{ref:Wang06}
\begin{equation}
R=1.70\pm0.03.
\end{equation}
From CMB constraint, the best fit values of parameters in dark
energy models can be determined by minimizing
\begin{equation}
\chi^{2}_{CMB}(p_s)=\frac{(R(p_s)-1.70)^2}{0.03^2}.\label{eq:chi2CMB}
\end{equation}

\begin{figure}[tbh]
\centering
\includegraphics[width=4.0in]{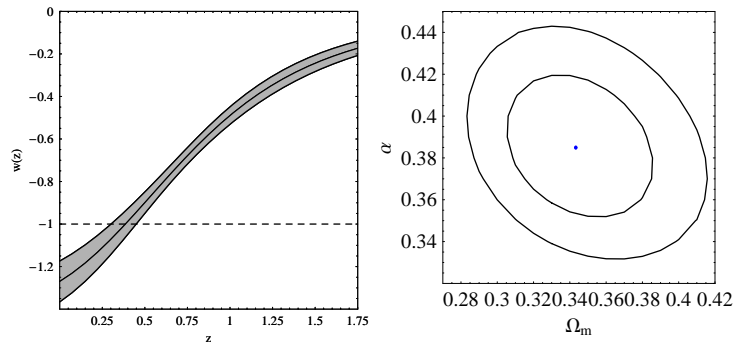}
\caption{The evolutions of EoS of RDE (the left panel) with respect
to the redshift $z$ and contour (the right panel) of $\Omega_{m0}$
and $\alpha$, where the parameters $\Omega_{m0}$ and $\alpha$ are
determined from SN+BAO+CMB. The center solid line is plotted with
the best fit values, where the shadows denote the $1\sigma$
regions.}\label{fig:wcz}
\end{figure}

For Gaussian distributed measurements, the likelihood function
$L\propto e^{-\chi^2/2}$, where $\chi^2$ is
\begin{equation}
\chi^2=\chi^2_{SNIa}+\chi^2_{BAO}+\chi^2_{CMB},
\end{equation}
here $\chi^2_{SNIa}$ is given in Eq. (\ref{eq:chi2SNIa}),
$\chi^2_{BAO}$ is given in Eq. (\ref{eq:chi2BAO}), $\chi^2_{CMB}$ is
given in Eq. (\ref{eq:chi2CMB}). Following the process described
above, the maximum likelihood corresponds to the minimum of $\chi^2$. We list the calculation results in Tab. \ref{tab:result}.
\begin{table}[tbh]
\begin{center}
\begin{tabular}{c|c|c|c}
\hline\hline Datasets & $\chi^2_{min}$ & $\Omega_{m0}$ & $\alpha$ \\
\hline SN+BAO+CMB & $202.25$  & $0.34\pm 0.04$ & $0.38\pm 0.03$  \\
\hline
\end{tabular}
\caption{The values of minimum $\chi^2$ and best fit values of the
parameters.}\label{tab:result}
\end{center}
\end{table}
Also, the evolutions of EOS of RDE and the contour with $1\sigma$
and $2\sigma$ error regions are plotted in Fig. \ref{fig:wcz}.

From Tab. \ref{tab:result}. and Fig. \ref{fig:wcz}, one can find
that the EoS of RDE is less than $-1$ at present with $1\sigma$ error. In the past, i.e.
at high redshift, the EoS approaches zero. So, RDE behaves like cold
dark matter in the past and like phantom at present. So, it is a
quintom like dark energy model.

In summary, in this paper, the viability of Ricci dark energy is
tested by confronting it with current cosmic observations including
type Ia supernovae (SN Ia), baryon acoustic oscillations (BAO) and
cosmic microwave background (CMB). By the maximum likelihood method,
we find the minimum $\chi^2_{min}=202.25$ and the best fit
parameters: $\Omega_{m0}=0.34\pm 0.04$, $\alpha=0.38\pm 0.03$ with
$1\sigma$ error. With these best fit values of the parameters, one
found out that Ricci dark energy is a viable dark energy model.

\acknowledgements{This work is supported by NSF (10573003), NSF
(10703001), NSF(10747113), SRFDP (20070141034) and NBRP
(2003CB716300) of P.R. China.}

\end{document}